\documentclass[aps,prb,twocolumn,groupedaddress]{revtex4-1}

\usepackage[pdftex,colorlinks,bookmarks = true]{hyperref} 
\usepackage{amsmath}
\usepackage{amsfonts}
\usepackage{amssymb}
\usepackage{empheq} 
\usepackage{mathtools} 
\usepackage{siunitx} 
\usepackage{csquotes} 
\usepackage{upgreek} 
\usepackage{braket} 
\usepackage[capitalize]{cleveref} 
\usepackage{subfig} 

\newcommand{\abs}[1]{\left| #1 \right|} 
\newcommand{\avg}[1]{\left\langle #1 \right\rangle} 
\renewcommand{\v}[1]{\boldsymbol{\mathbf{#1}}} 
\newcommand{\inv}{^{-1}} 
\newcommand{\e}[1]{\, \mathrm{e}^{#1}} 
\renewcommand{\i}{\mathrm{i}} 
\newcommand{\cpi}{\uppi} 
\providecommand*{\CP}[0]{{\mathbb{CP}^1}} 
\providecommand*{\groupU}[1]{\mathrm{U}(#1)} 
\providecommand*{\groupO}[1]{\mathrm{O}(#1)} 

\makeatletter 
\providecommand*{\diff}
	  {\@ifnextchar^{\DIfF}{\DIfF^{}}}
  \def\DIfF^#1{%
	  \mathop{\mathrm{\mathstrut d}}%
		  \nolimits^{#1}\gobblespace}
  \def\gobblespace{%
	  \futurelet\diffarg\opspace}
  \def\opspace{%
	  \let\DiffSpace\!%
	  \ifx\diffarg(%
		  \let\DiffSpace\relax
	  \else
		  \ifx\diffarg[%
			  \let\DiffSpace\relax
		  \else
			  \ifx\diffarg\{%
				  \let\DiffSpace\relax
			  \fi\fi\fi\DiffSpace}
			  
\makeatother
\providecommand*{\deriv}[3][]{\frac{\diff^{#1}#2}{\diff #3^{#1}}} 

\begin{document}

\title{Berry phases, current lattices, and suppression of phase transitions in a lattice gauge theory of quantum antiferromagnets}

\author{Troels Arnfred Bojesen}
\email[]{troels.bojesen@ntnu.no}
\affiliation{Department of Physics, Norwegian University of Science and Technology, N-7491 Trondheim, Norway}

\author{Asle Sudb\o{}}
\email[]{asle.sudbo@ntnu.no}
\affiliation{Department of Physics, Norwegian University of Science and Technology, N-7491 Trondheim, Norway}

\date{\today}

\begin{abstract}
We consider a lattice model of two complex scalar matter fields $z_{a}, a=1,2$ under a $\CP$-constraint $\abs{z_1}^2+\abs{z_2}^2=1$, minimally coupled to a compact gauge 
field, with an additional Berry phase term. This model has been the origin of a large body of works addressing novel paradigms for quantum criticality,
in particular \enquote{spin-quark} (spinon) deconfinement in $S=1/2$ quantum antiferromagnets. We map the model exactly to a link-current model, which permits the use of 
classical worm algorithms to study the model in large-scale Monte Carlo simulations on lattices of size $L^3$, up to $L=512$. We show that the addition of a Berry 
phase term to the lattice $\CP$-model completely suppresses the phase transition in the $\groupO{3}$ universality class of the $\CP$-model, such that the original spin-system
described by the compact gauge theory is always in the ordered phase. The link-current formulation of  the model is useful in identifying the mechanism by which the phase 
transition from an ordered to a disordered state is suppressed. 
\end{abstract}

\maketitle

\section{Introduction}

Models of several complex scalar matter fields minimally coupled to compact and non-compact gauge fields, have been intensively studied in 
condensed matter physics over the last decade.\cite{Bernevig_AP_2004,Senthil_Science_2004,Senthil_PRB_2004,Motrunich_PRB_2004,Kuklov_Ann_Phys_2006,Smiseth_PRB_2005,Kragset_PRL_2006,Motrunich_ArXiv_2008,Kuklov_PRL_2008,Dahl_PRB_2008,Bonderson_PRB_2011,Herland_PRB_2012} The main motivation for this has been that these models appear to find their realization in quite disparate condensed matter systems. Examples are 
low-dimensional quantum spin system emerging as effective low-energy descriptions of insulating phases of Mott-Hubbard insulators,\cite{Senthil_Science_2004,Senthil_PRB_2004} 
multicomponent superconductors and superfluids,
\cite{Kuklov_Ann_Phys_2006,Smiseth_PRB_2005,Kragset_PRL_2006,Motrunich_ArXiv_2008,Kuklov_PRL_2008,Dahl_PRB_2008,Bonderson_PRB_2011,Herland_PRB_2012} 
and plasma analogs of the norms of non-Abelian fractional quantum Hall states.\cite{Bonderson_PRB_2011,Herland_PRB_2012} Of particular interest has been 
the issue of whether or not such theories feature phase-transitions which are unconventional in the sense of being difficult to describe within a 
Landau-Wilson-Ginzburg paradigm of phase transitions.\cite{Senthil_Science_2004,Senthil_PRB_2004,Kuklov_Ann_Phys_2006,Kragset_PRL_2006} The notion of 
deconfined quantum criticality,\cite{Senthil_Science_2004,Senthil_PRB_2004} whereby a phase transition takes place by deconfinement of basic building 
blocks (spinons) for various ordering fields, rather than through the standard mechanism of spontaneous symmetry breaking, has been central in this context.  

A key step in many of the investigations of $T=0$ phase transitions in quantum spin models is to rewrite the spin-operator in terms of two
complex scalar matter fields $z_{j a}, a=1,2$, namely $\v{S}_j = S \v{n}_j$, where $S$ is the length of the spin and $\v{n}_j$ is a unit vector 
living on the $2$-sphere, given by 
\begin{equation}
 \v{n}_j = z^*_{j a} \v{\sigma}_{a b} z_{j \beta}.
 \label{eq:z_definition}
\end{equation}
The local constraint $\abs{\v{n}_j}=1$ translates into the $\CP$-constraint
$\abs{z_{j1}}^2+\abs{z_{j2}}^2=1$. The above \enquote{spinon}-representation of the spin-operator immediately introduces a gauge-symmetry in the problem, 
since  $\v{S}_j$ is invariant under the local transformation   $z_{j a} \to z_{j a} \e{\i \phi_j}$. The associated $\groupU{1}$ gauge field is compact, 
defined modulo  $2 \cpi$. Based on the above, one arrives at the following gauge-theory action of a quantum spin system  on a bipartite 
lattice\cite{Sachdev_Tutorial,Sachdev_Book} 
\begin{equation}
\mathcal{S} = \frac{1}{g} \sum_{\langle i,j \rangle} \v{n}_i \cdot \v{n}_j + \i 2 S \sum_j \eta_j \mathcal{A}_{j \tau}. 
\label{eq:Action_1}
\end{equation}
Here, $1/g$ is a measure of the nearest neighbor spin-coupling in the problem, $\eta_j = \pm 1$ is a staggering factor whose sign depends on which 
sublattice the spin is located. The second term is the Berry-phase term that one obtains in a functional integral formulation, where $\mathcal{A}_{j \tau}$ 
is a local portion of the closed curve enclosing the areas subtended by a fluctuating quantum spin as the system evolves in imaginary time from $\tau=0$ 
to $\tau=\beta$, where $\beta$ is the inverse temperature. We will consider the system in the limit $\beta \to \infty$. The local portion of the curve is 
taken between neighboring sites in the imaginary time direction, once this direction has been discretized. Note that both $\v{n}_i$ and $\mathcal{A}_{j \tau}$ 
depend on the spinon fields $z_{j \alpha}$, something that makes calculations quite awkward. A reformulation of the model such that it is expressed in terms 
of the spinon fields and an independently fluctuating gauge field was proposed in Ref. \onlinecite{Sachdev_Jalabert_1990}, and it is this version of the 
model we will consider in the present paper. It is essentially a lattice $\CP$-model augmented by a term mimicking the imaginary Berry-phase term 
in \cref{eq:Action_1}. This term will have a decisive influence on the phase transition of the model.

\section{The Model and mapping}
The model we will consider in this paper is given by \cite{Sachdev_Tutorial,Sachdev_Book}
\begin{multline}
 \mathcal{Z} = \prod_{j\mu}\int_{0}^{2\cpi}\frac{\diff A_{j\mu}}{2\cpi} \prod_{ja} \int \diff z_{ja} \diff z_{ja}^{*} \\
 \exp\left[g\inv \sum_{ja\mu} \left(z_{ja}^{*}\e{-\i A_{j\mu}}z_{j+\mu,a} + \text{c.c.}\right) + \i 2S \sum_j \eta_j A_{j\tau} \right],
 \label{eq:Z_SJ}
\end{multline}
with the local $\CP$-constraint
\begin{equation}
\abs{z_{j1}}^{2} + \abs{z_{j2}}^{2} = 1 \quad \forall j.
\label{eq:CP1_constraint}
\end{equation}
The lattice is cubic, and we use $\mu \in \set{x,y,\tau}$ as a (positive) direction index, as well as a unit 
vector in that direction. The meaning should be clear from the context. The scalar matter fields $z_{ja}$ live on the lattice sites, 
while $A_{j\mu}$ is a $z$-\emph{independent} $\groupU{1}$ gauge field living on the links $(j,j+\mu)$. $\eta_j$ is a Neel staggering 
factor being $+1(-1)$ on spatial sublattice A(B). The first term of the action resembles that of a lattice $\CP$-model, while the second is 
an additional Berry-phase term. The connection between Eqs. \ref{eq:Action_1} and \ref{eq:Z_SJ} is given in \onlinecite{Sachdev_Tutorial,Sachdev_Book}.

Note the absence of any Maxwell-like term in the above model. Several previous treatments of the  problem have added a Maxwell-term, either compact or 
non-compact, to the action. The rationale for doing this is that  integrating out the Fourier-components of the matter field at large momenta (short-distance 
physics), must yield a term involving only the gauge-field. Since the term needs to be gauge-invariant, a non-compact or compact Maxwell term is often written 
down.  We will refrain from this in the present paper, since there appears no Mawell-like term of the gauge-field in the basic action \cref{eq:Action_1}, and 
therefore not in \cref{eq:Z_SJ}. Moreover, the Monte Carlo procedure integrates out the short-distance physics of the matter-field in the problem, so if a 
Maxwell-like term is generated dynamically \cite{Sachdev_Jalabert_1990}, it will be implicitly included in the description. We emphasize that the conclusions we 
draw in this paper are based on simulations of the model Eq. 3 with the constraint Eq. 4. It may be that a gauge-theory formulation of a more general 
microscopic spin model than Eq. 2 model will feature different results, see also comments below.  

Due to the imaginary Berry-phase term, direct simulation of this model is technically difficult. However, a major advance on the problem
can be made by mapping it \emph{exactly} onto a real-valued link-current (LC) model, which in turn can be efficiently dealt with 
using a worm algorithm \cite{Prokof'ev_Svistunov_2001}. Details of this mapping is given in \cref{A:current_mapping}. The mapping also 
obviates the need that arises of introducing, by hand, a Maxwell-term in order to regularize the functional integrations in a direct 
representation. The result reads
\begin{equation}
 \mathcal{Z} = \sum_{\set{J}} \prod_{ja\kappa}\frac{g^{-J_{ja\kappa}}}{J_{ja\kappa}!} \prod_{j}\frac{\mathcal{N}_{j1}!\mathcal{N}_{j2}!}{\left(\mathcal{N}_{j1} 
+ \mathcal{N}_{j2} + 1\right)!},
 \label{eq:Z_current}
\end{equation}
with the constraints
\begin{gather}
 \sum_{\kappa}I_{ja\kappa} = 0, \label{eq:kirchhoff} \\
 I_{j1x} + I_{j2x} = 0, \label{eq:I_x_constraint} \\
 I_{j1y} + I_{j2y} = 0, \label{eq:I_y_constraint}\\
 I_{j1\tau} + I_{j2\tau} + 2 S \eta_j = 0 \label{eq:I_tau_constraint},
\end{gather}
where 
\begin{equation}
  I_{ja\kappa} \equiv J_{ja\kappa} - J_{j+\kappa,a,-\kappa} = -I_{j+\kappa,a,-\kappa} \in \mathbb{Z}.
\end{equation}
Moreover, $J_{ja\kappa} \in \mathbb{N}_{0}$ denotes the \emph{non-negative} integer current of component $a$ on the link 
going \emph{from} lattice site $j$ \emph{to} a neighboring lattice site $j+\kappa$, with $\kappa \in \set{\pm x, \pm y, \pm \tau}$. Note that, contrary to 
the $I$-current, the $J$-current going in the opposite direction on the $(j,j+\kappa)$-link is another degree of freedom; generally $J_{ja\kappa} \neq J_{j+\kappa,a,-\kappa}$. Finally, we have introduced $\mathcal{N}_{ja} = \sum_{\kappa} J_{ja\kappa}$ and the notation $\set{J}$ for the set of all possible, permissible current field configurations. In our simulations, we set $S=1/2$.  

It will turn out that the term $2 S \eta_j$ on the left hand side of Eq. \ref{eq:I_tau_constraint} will play a crucial role in the following. If we view the quantities
$I_{j \alpha \tau}$ as currents in the imaginary-time direction on the space-time lattice, the case $\eta_j=0$ corresponds to the case where there is no imposed 
background current lattice in the $\tau$-direction. The phase transition of the model then proceeds via a current-loop blowout in the background of zero current lattice,
and will be discussed in detail below. This transition has a well-known analogy, namely the phase-transition from a superconductor to a normal metal via a vortex-loop blowout 
in a type-II superconductor in zero magnetic field. However, as soon as $\eta_j \neq 0$, i.e. when a 
Berry-phase staggering factor is introduced, the situation is drastically altered.  Now, there {\it is} a background current-lattice imposed on the system in 
the $\tau$-direction. This, it will turn out, suffices to destroy the phase transition in much the same way as a vortex-loop blowout transition may be suppressed 
by the presence of a vortex-lattice in a type-II superconductor, see Sections III and V for a more detailed discussion.  

It is also worth noting how different the loop-current model given above, starting from \cref{eq:Z_SJ}, is compared to what we would find were we to add a 
Maxwell-term right from the start in \cref{eq:Z_SJ}. In the latter case, there would be no constraints \cref{eq:I_x_constraint,eq:I_y_constraint,eq:I_tau_constraint}. 
Instead, these constraints would be replaced by long-range interactions between current-segments living on the links of the space-time lattice \cite{Herland_PRB_2012}.  
A loop-model formulation of the $\CP$-model (i.e. with no Berry-phase term) has previously been provided in Ref. \onlinecite{Wolff_NPB_2010}.

The observable we choose to study is the winding number in the $\mu$-direction, given by
\begin{equation}
 W_{\mu} \equiv \frac{1}{L_{\mu}}\sum_j I_{j1\mu} = -\frac{1}{L_{\mu}}\sum_j I_{j2\mu},
\end{equation}
where $L_{\mu}$ is the system size in the $\mu$ direction. The winding number of the LC model is related to the gauge invariant phase stiffness of the original 
model, \cref{eq:Z_SJ}, which in term of the link-currents is given by
\begin{equation}
 \varUpsilon_{\mu} = \left.\frac{1}{L_\mu}\deriv[2]{\ln \mathcal{Z}'}{\delta_{\mu}}\right|_{\delta_{\mu} = 0} = \frac{1}{L_{\mu}}\avg{W_\mu^2}.
 \label{eq:stiffness}
\end{equation}
Here, the primed $\mathcal{Z}$ indicates that we have introduced a gauge invariant phase twist 
$(\theta_{j1},\theta_{j2}) \to (\theta_{j1},\theta_{j2}) + (\v \delta \cdot \v r(j),-\v \delta \cdot \v r(j))$ in the phases of $ z_{ja} \sim \e{\i \theta_{ja}}$. $\v r(j)$ 
is the coordinate vector $\v r$ at lattice site $j$, and $\delta_{\mu}$ is the $\mu$ component of $\v \delta$. 
We expect the stiffness to scale as $\varUpsilon_{\mu} \sim L_{\mu}^{2-d} = L_{\mu}^{-1}$ at criticality, while $\varUpsilon_{\mu} \sim \mathcal{O}(1)$ in 
the ordered phase.\cite{Sandvik_2010} Hence we have that $\avg{W_{\mu}^2} = \mathcal{O}(1)$ at criticality, and we can use the scale invariant crossing point 
of $\avg{W_{\mu}^2}$-curves to determine the critical point, if there indeed is one. 

In the simulations, we have chosen $L_x = L_y = L_\tau = L$ and computed the average of the winding number in the $x$ and $y$ direction,
\begin{equation}
 W_{xy}^2 \equiv \frac{1}{2}\left(W_x^2 + W_y^2\right).
\end{equation}

This suffices to investigate spatial spin-ordering, which is the relevant component of the winding number when considering the competition between Neel order and 
the emergence of a valence bond solid.  

\section{Simulation Technique}
Link-current models can be efficiently simulated using worm algorithms.\cite{Prokof'ev_Svistunov_2009} The most efficient worm algorithms at the moment are, to our 
knowledge, the geometrical worm algorithms.\cite{Fabien_Sorensen_undir_2003, Fabien_Sorensen_dir_2003} However, the great number of local degrees of freedom (24) 
when moving the worm through the lattice of the LC model, together with the lattice site coupling factors $\mathcal{N}_{ja}$, renders a geometrical approach 
too memory-demanding. Hence, a \enquote{classical} worm algorithm \cite{Prokof'ev_Svistunov_2001} was chosen.

The non-negativity of the $J$-currents means that extra care must be taken when the head of the \enquote{worm} is updated. To fulfill \enquote{Kirchhoff's law}, 
\cref{eq:kirchhoff}, the $I_{ja\kappa}$-current is updated with +1 if the new site is in the positive lattice direction, and -1 if the new site is in the 
negative lattice direction. This requirement can be met in two ways for each component. Namely, one can either have 
$J_{ja\kappa} \gets J_{ja\kappa} \pm 1$, or $J_{j+\kappa,a,-\kappa} \gets J_{j+\kappa,a,-\kappa} \mp 1$. The two possibilities are chosen with equal 
probability at each proposed move, with the extra constraint that only the $J \gets J + 1$ update can be chosen if $J = 0$. This constraint does 
not alter the probability distribution of the updates of the two matter-field components, as the situation is symmetric with respect to $a = 1 \leftrightarrow 2$.

If we disregard the staggering factor $\eta_j$ for a moment, the effect of \cref{eq:I_x_constraint,eq:I_y_constraint,eq:I_tau_constraint} is basically 
that both components share the same \enquote{worm}, and are updated at the same time, but with opposite signs. In total, we may therefore have  
4 possible $J$-current update \enquote{routes} when moving the head of the worm to a neighboring site.

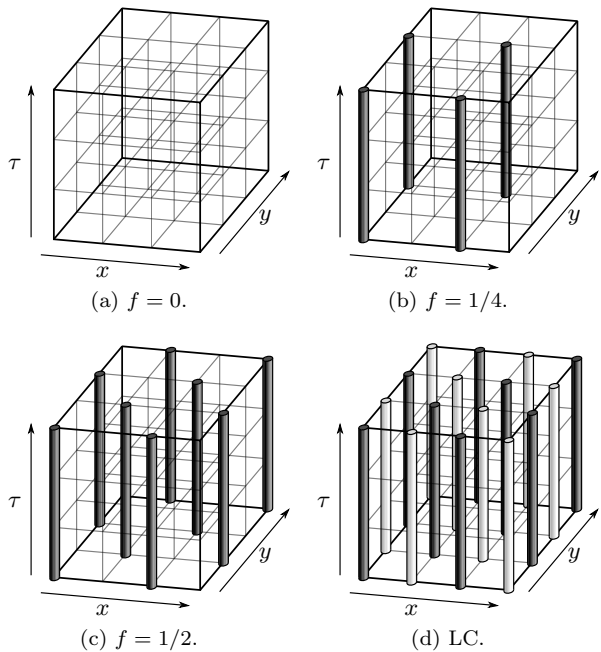
\begin{figure}
\def\svgwidth{0.45\columnwidth}
\subfloat[$f=0$.]{
\begingroup%
  \makeatletter%
  \providecommand\color[2][]{%
    \errmessage{(Inkscape) Color is used for the text in Inkscape, but the package 'color.sty' is not loaded}%
    \renewcommand\color[2][]{}%
  }%
  \providecommand\transparent[1]{%
    \errmessage{(Inkscape) Transparency is used (non-zero) for the text in Inkscape, but the package 'transparent.sty' is not loaded}%
    \renewcommand\transparent[1]{}%
  }%
  \providecommand\rotatebox[2]{#2}%
  \ifx\svgwidth\undefined%
    \setlength{\unitlength}{199.46101588bp}%
    \ifx\svgscale\undefined%
      \relax%
    \else%
      \setlength{\unitlength}{\unitlength * \real{\svgscale}}%
    \fi%
  \else%
    \setlength{\unitlength}{\svgwidth}%
  \fi%
  \global\let\svgwidth\undefined%
  \global\let\svgscale\undefined%
  \makeatother%
  \begin{picture}(1,0.92680039)%
    \put(0,0){\includegraphics[width=\unitlength]{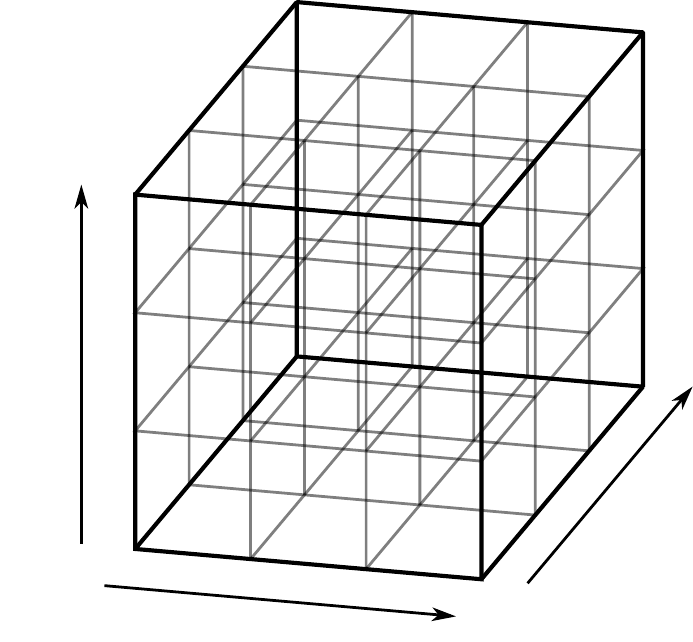}}%
    \put(0.36183749,0.00589479){\color[rgb]{0,0,0}\makebox(0,0)[b]{\smash{$x$}}}%
    \put(0.91730412,0.19889391){\color[rgb]{0,0,0}\makebox(0,0)[b]{\smash{$y$}}}%
    \put(0.06188553,0.37321403){\color[rgb]{0,0,0}\makebox(0,0)[b]{\smash{$\tau$}}}%
  \end{picture}%
\endgroup%
}
\def\svgwidth{0.45\columnwidth}
\subfloat[$f=1/4$.]{
\begingroup%
  \makeatletter%
  \providecommand\color[2][]{%
    \errmessage{(Inkscape) Color is used for the text in Inkscape, but the package 'color.sty' is not loaded}%
    \renewcommand\color[2][]{}%
  }%
  \providecommand\transparent[1]{%
    \errmessage{(Inkscape) Transparency is used (non-zero) for the text in Inkscape, but the package 'transparent.sty' is not loaded}%
    \renewcommand\transparent[1]{}%
  }%
  \providecommand\rotatebox[2]{#2}%
  \ifx\svgwidth\undefined%
    \setlength{\unitlength}{199.46101588bp}%
    \ifx\svgscale\undefined%
      \relax%
    \else%
      \setlength{\unitlength}{\unitlength * \real{\svgscale}}%
    \fi%
  \else%
    \setlength{\unitlength}{\svgwidth}%
  \fi%
  \global\let\svgwidth\undefined%
  \global\let\svgscale\undefined%
  \makeatother%
  \begin{picture}(1,0.92680039)%
    \put(0,0){\includegraphics[width=\unitlength]{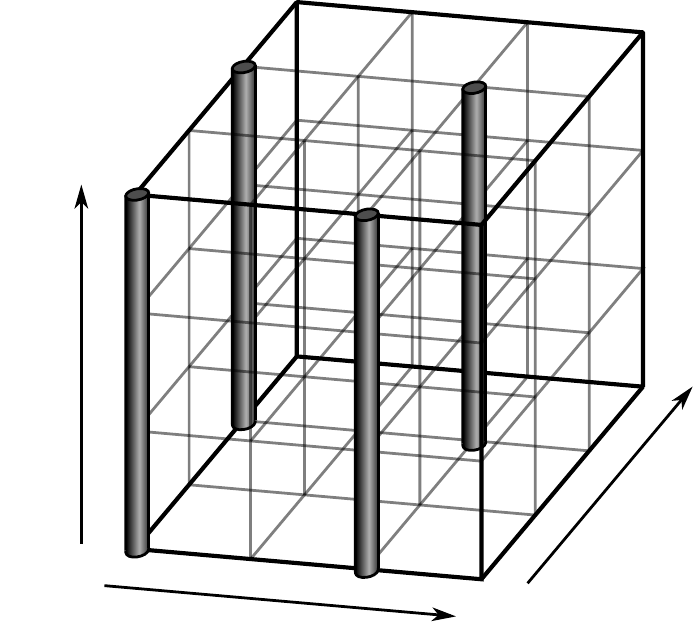}}%
    \put(0.36183749,0.00589479){\color[rgb]{0,0,0}\makebox(0,0)[b]{\smash{$x$}}}%
    \put(0.91730412,0.19889391){\color[rgb]{0,0,0}\makebox(0,0)[b]{\smash{$y$}}}%
    \put(0.06188553,0.37321403){\color[rgb]{0,0,0}\makebox(0,0)[b]{\smash{$\tau$}}}%
  \end{picture}%
\endgroup%
}
\def\svgwidth{0.45\columnwidth}
\subfloat[$f=1/2$.]{
\begingroup%
  \makeatletter%
  \providecommand\color[2][]{%
    \errmessage{(Inkscape) Color is used for the text in Inkscape, but the package 'color.sty' is not loaded}%
    \renewcommand\color[2][]{}%
  }%
  \providecommand\transparent[1]{%
    \errmessage{(Inkscape) Transparency is used (non-zero) for the text in Inkscape, but the package 'transparent.sty' is not loaded}%
    \renewcommand\transparent[1]{}%
  }%
  \providecommand\rotatebox[2]{#2}%
  \ifx\svgwidth\undefined%
    \setlength{\unitlength}{199.46101588bp}%
    \ifx\svgscale\undefined%
      \relax%
    \else%
      \setlength{\unitlength}{\unitlength * \real{\svgscale}}%
    \fi%
  \else%
    \setlength{\unitlength}{\svgwidth}%
  \fi%
  \global\let\svgwidth\undefined%
  \global\let\svgscale\undefined%
  \makeatother%
  \begin{picture}(1,0.92680039)%
    \put(0,0){\includegraphics[width=\unitlength]{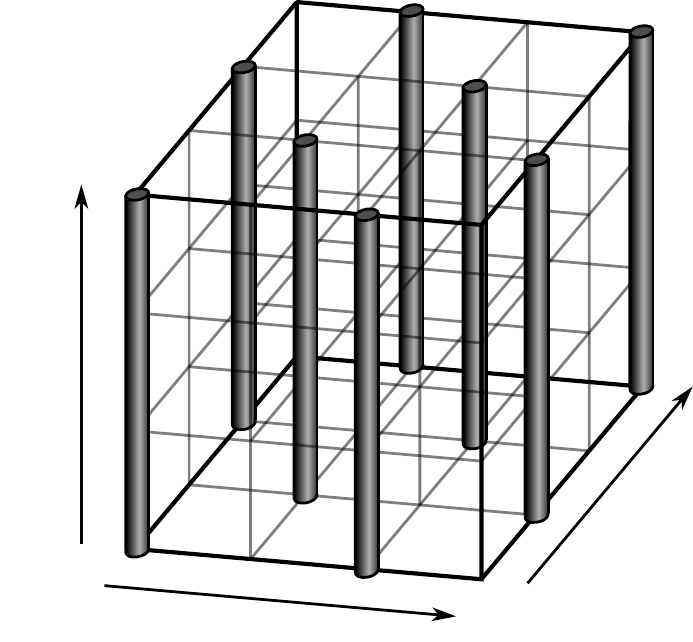}}%
    \put(0.36183749,0.00589479){\color[rgb]{0,0,0}\makebox(0,0)[b]{\smash{$x$}}}%
    \put(0.91730412,0.19889391){\color[rgb]{0,0,0}\makebox(0,0)[b]{\smash{$y$}}}%
    \put(0.06188553,0.37321403){\color[rgb]{0,0,0}\makebox(0,0)[b]{\smash{$\tau$}}}%
  \end{picture}%
\endgroup%
}
\def\svgwidth{0.45\columnwidth}
\subfloat[LC.\label{fig:SJ_background}]{
\begingroup%
  \makeatletter%
  \providecommand\color[2][]{%
    \errmessage{(Inkscape) Color is used for the text in Inkscape, but the package 'color.sty' is not loaded}%
    \renewcommand\color[2][]{}%
  }%
  \providecommand\transparent[1]{%
    \errmessage{(Inkscape) Transparency is used (non-zero) for the text in Inkscape, but the package 'transparent.sty' is not loaded}%
    \renewcommand\transparent[1]{}%
  }%
  \providecommand\rotatebox[2]{#2}%
  \ifx\svgwidth\undefined%
    \setlength{\unitlength}{199.46101588bp}%
    \ifx\svgscale\undefined%
      \relax%
    \else%
      \setlength{\unitlength}{\unitlength * \real{\svgscale}}%
    \fi%
  \else%
    \setlength{\unitlength}{\svgwidth}%
  \fi%
  \global\let\svgwidth\undefined%
  \global\let\svgscale\undefined%
  \makeatother%
  \begin{picture}(1,0.93438539)%
    \put(0,0){\includegraphics[width=\unitlength]{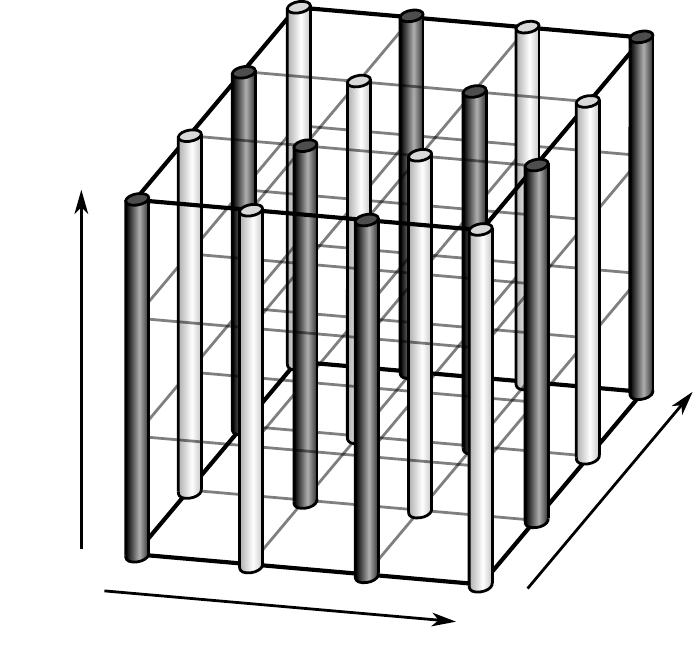}}%
    \put(0.36183749,0.00589479){\color[rgb]{0,0,0}\makebox(0,0)[b]{\smash{$x$}}}%
    \put(0.91730412,0.19889391){\color[rgb]{0,0,0}\makebox(0,0)[b]{\smash{$y$}}}%
    \put(0.06188553,0.37321403){\color[rgb]{0,0,0}\makebox(0,0)[b]{\smash{$\tau$}}}%
  \end{picture}%
\endgroup%
}
\caption{Some background current field examples for a $L=4$ lattice. Dark and light cylinders represent $I_{\tau}=1$ and $I_{\tau}=-1$, respectively. \label{fig:background}}
\end{figure}

The fixed staggering field $\eta$ can easily be dealt with if we treat it as a background staggering current field, as illustrated in \cref{fig:SJ_background}. 
If we initialize the $J$-current field such that (for instance) $I_{A1\tau} = -1$ and $I_{B1\tau} = 1$ and $I = 0$ for all other currents, we can treat 
worm moves in all directions in the same way, as explained above, and \cref{eq:I_tau_constraint} will still be fulfilled when the worm forms a closed 
loop. Thus, one way of viewing the effect of the Berry-phase term in the link-current representation, is that the link-currents $I_{j a \tau}$, which 
fluctuate in a vacuum in the standard $\CP$-model, instead fluctuate in the background of a current lattice when the Berry-phase is introduced. This has 
some resemblance to the vortex-loop blowout that drives the superfluid-normal fluid phase transition, or the superconductor-normal metal phase transition 
in a type-II superconductor. The standard $\CP$-model corresponds roughly to the absence of rotation or magnetic field in the superfluid or superconductor, 
respectively, while the presence of the Berry-phase term corresponds to the presence of rotation or magnetic field, see Appendix \ref{A:current_mapping} 
for details.

By using the identity $\exp \ln x = x$ on the summand of \cref{eq:Z_current}, it is easy to see that the LC model can be written on a form resembling a 
partition function of the canonical ensemble, with an analogous inverse temperature $\beta = \ln g$. Doing this has made it possible to use standard 
Ferrenberg-Swendsen multi-histogram reweighting \cite{Ferrenberg_PRL_1989} to improve our numerical data.

Pseudorandom numbers were generated by the Mersenne-Twister algorithm.\cite{Matsumoto_1998_ACM_TMCS} Errors were determined using the jackknife method.

\section{Berry-phase suppression of the phase transition in the LC model}
We claim that in the presence of the Berry-phase term in \cref{eq:Z_SJ}, there is \emph{no} phase transition in the LC model, and that it is the staggering 
field $\eta$ which is responsible for this. We show this by starting with the $\CP$ lattice model (i.e. the LC model without the Berry-phase term), 
which \emph{has} a phase transition,\cite{Takashima_PRB_2005} (see also Appendix \ref{C:exponents}) and gradually increase a background current field 
in the $\tau$ direction. We show that even a weak background field destroys the phase transition, and this happens regardless of whether the background 
field is staggered or not.

\begin{figure}[tbp]
  \includegraphics{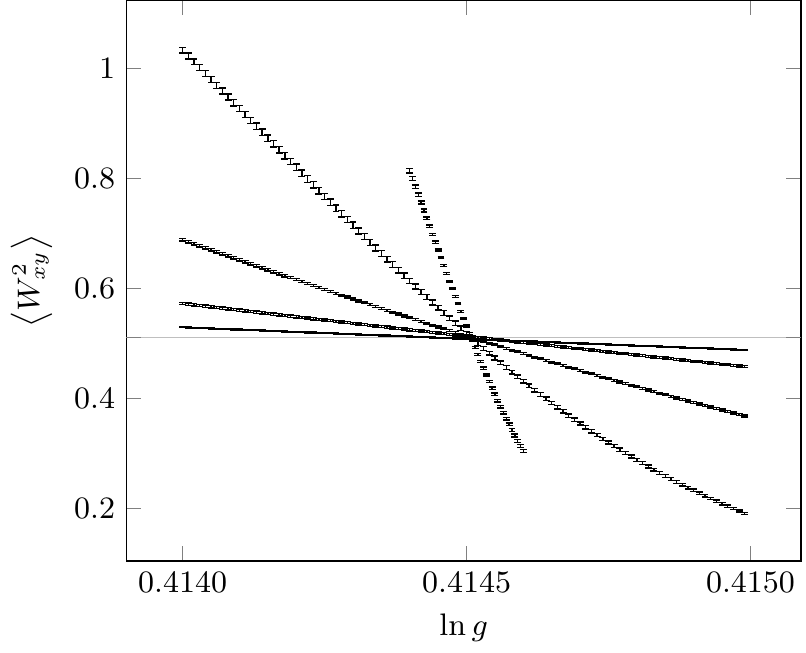}%
  \caption{Finite size scaling for $\avg{W_{xy}^2}$-curves for the $\CP$-model. $L=16,32,64,128,256$. The horizontal line at $\avg{W_{xy}^2} = 0.511$ indicates 
the (approximate) size independent crossing point. }
  \label{fig:w2_CP1}
\end{figure}

\begin{figure}[tbp]
  \includegraphics{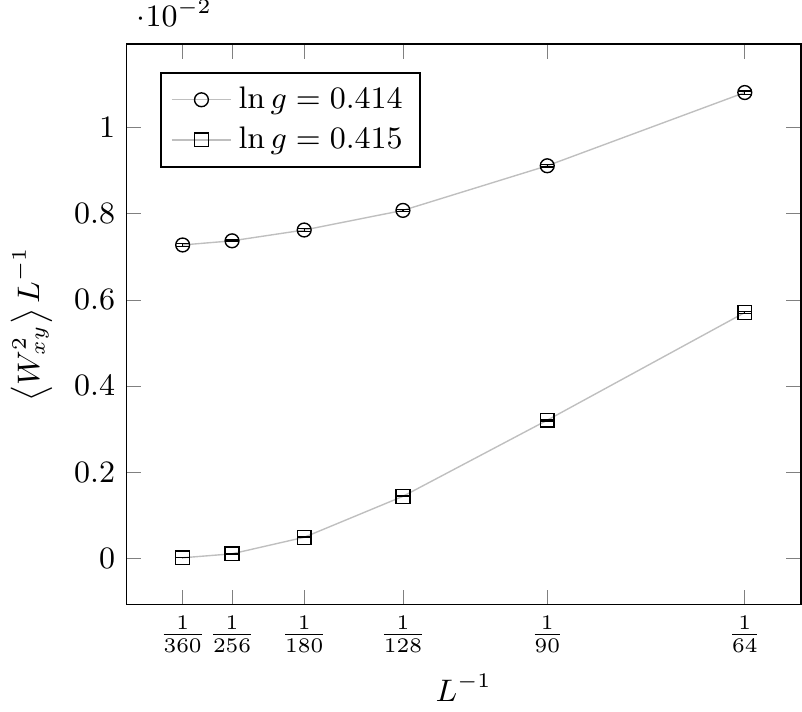}%
  \caption{Finite size scaling for the phase stiffness $\avg{W_{xy}^2}L\inv$ for the $\CP$-model, showing the behavior on each side of the critical point 
$\ln g_\text{c} \approx 0.4145$. Error bars are smaller than symbol sizes. Lines are guides to the eye.}
  \label{fig:rho_CP1}
\end{figure}

$\avg{W_{xy}^2}$-curves for the $\CP$-model are shown in \cref{fig:w2_CP1} for system sizes $L=16,32,64,128,256$. All the curves intersect at approximately 
the same point, $(\ln g_\text{c}, \avg{W_{xy}^2}) \approx (0.4145,0.511)$ -- as expected from finite size scaling (FSS) for a phase transition. The phase 
transition is verified in \cref{fig:rho_CP1}, where the phase stiffness $\avg{W_{xy}^2} L\inv$ is shown to go to a finite value for a coupling less than 
the critical coupling $\ln g_\text{c}$ and to zero for a coupling greater than $\ln g_\text{c}$.

\begin{figure}[tbp]
  \includegraphics{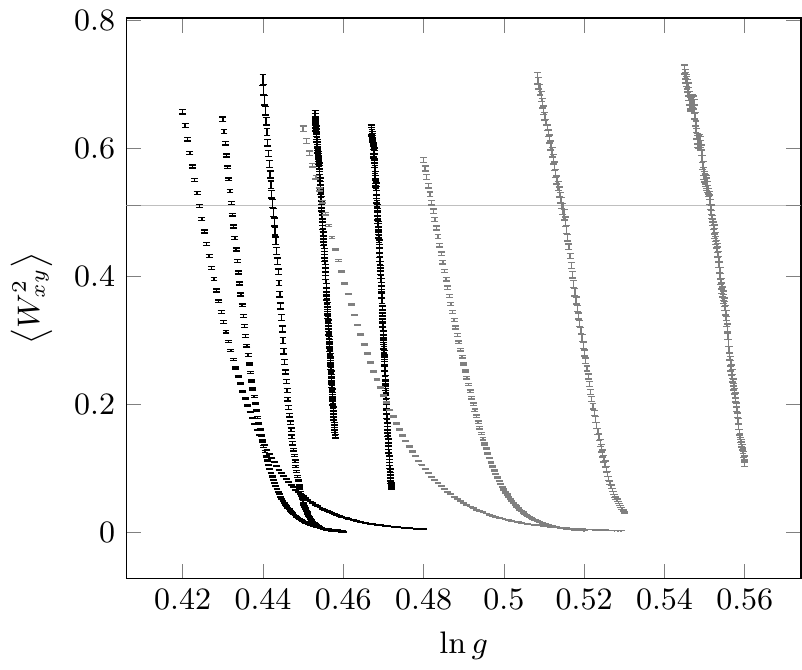}%
  \caption{Finite size scaling for $\avg{W_{xy,f}^2}$-curves for the $f = 1/64$-model (black) and the $f=1/16$-model (gray). $L=16,32,64,128,256$ for $f=1/64$ 
and $L=16,32,64,128$ for $f=1/16$. The horizontal line at $\avg{W_{xy}^2} = 0.511$ indicates the (approximate) size independent crossing point value of the 
$\CP$-model ($f=0$).}%
  \label{fig:w2_fill}%
\end{figure}

A background field is introduced by initializing a fraction $f$ of the lattice with a nonzero current $I_{j1\tau} = 1$, i.e. $f = L\inv \sum_j I_{j1\tau}$. 
See \cref{fig:background}. If such a field (with $f>0$) is included in the model, the situation changes dramatically. Let $\avg{W_{xy,f}^2}$ denote 
$\avg{W_{xy}^2}$ in this case. \Cref{fig:w2_fill} shows $\avg{W_{xy,f}^2}$-curves for $f=1/64$ and $f=1/16$, which is to be compared with \cref{fig:w2_CP1}. 
The curves shift to the right as the system size is increased, with no signs of converging even for large systems; there are no size-independent crossing 
points.

\begin{figure}[tbp]
  \includegraphics{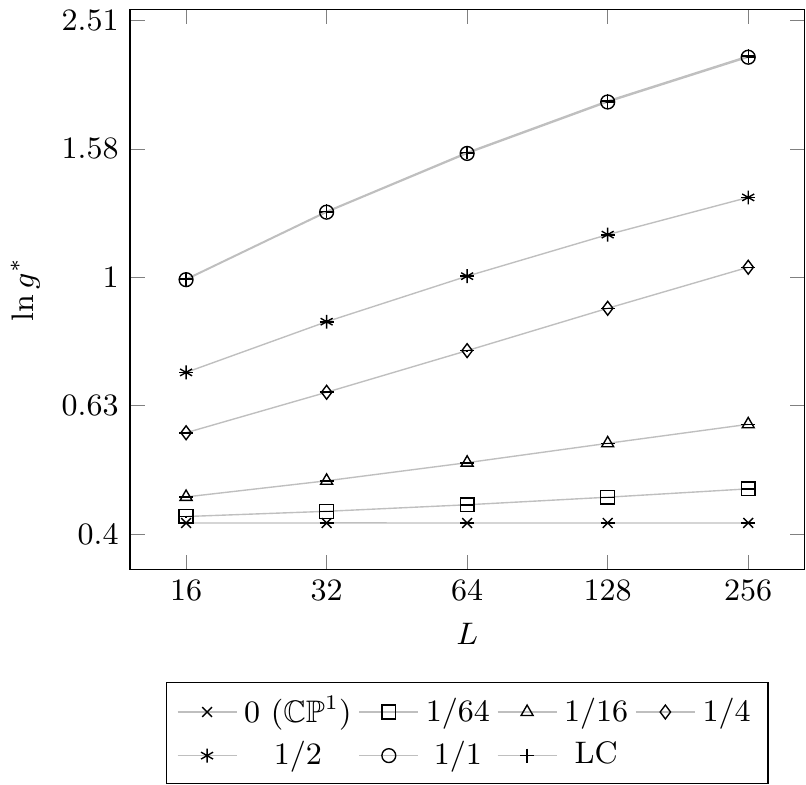}%
  \caption{Log-log plot of $\ln g^*_f$ as a function of system size $L$ for $f=0, \ldots,1$ as well as the LC model. The LC curve lies slightly above the $f=1$-curve. 
Error bars are smaller than symbol sizes. Lines are guides to the eye.}
  \label{fig:divergence}
\end{figure}

The divergence of the $\avg{W_{xy,f}^2}$-curves becomes clearer if we define a pseudocritical (finite-size critical) coupling $\ln g^*_{f}$ by choosing the 
value of  $\avg{W_{xy,f}^2}(\ln g^*_{f}) = 0.511 \approx \avg{W_{xy,0}^2}(\ln g_\text{c})$ for all $f$, and plot $\ln g^*_{f}$ as a function of system size. 
This is shown in \cref{fig:divergence} for system sizes $L=16,\ldots,256$ and several $f$-values up to 1 (maximal uniform background current field), in addition 
to the result for the LC model (maximally staggered background current field). It is seen from \cref{fig:divergence} that $\ln g^*_{f}$  increases monotonically 
with $L$ for all values of $f$, and more so for larger values of $f$ than for small values of $f$. The increase in  $\ln g^*_{f}$  is clearly seen also for $f=1/64$, 
which is the smallest value we have considered. The range of values that $\ln g^*_{f}$  display in a given interval of $L$-values, increases with $f$. For $\ln g^*_{f}$ 
to take on values spanning a decade for $f=1/64$ would require enormous system sizes, beyond the capability of present-day computers. For $f=1$, the case relevant for 
quantum antiferromagnets on a bipartite lattice, it is seen from \cref{fig:divergence} that the range of values is much larger in the interval $L \in [16,\ldots,256]$.

\begin{figure}[tbp]
  \includegraphics{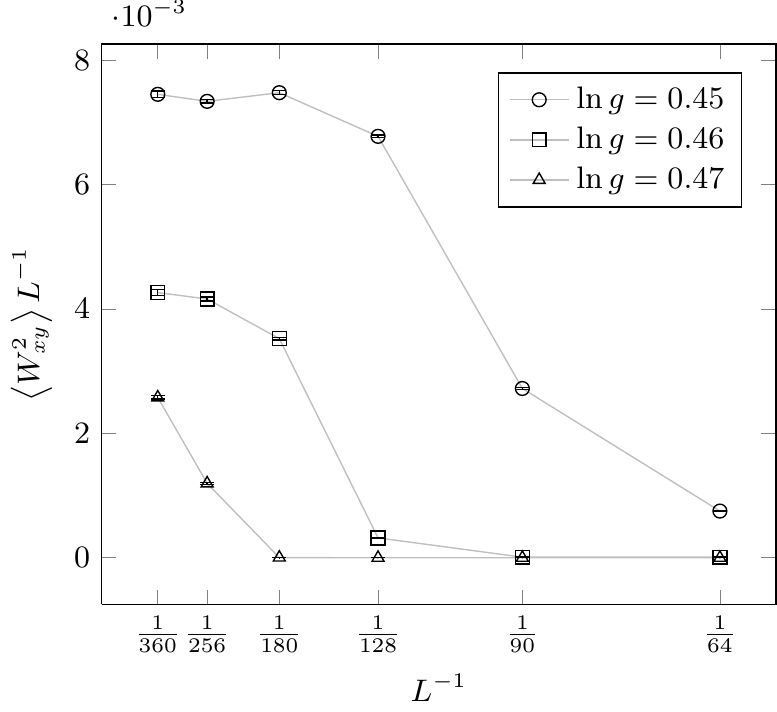}%
  \caption{Finite size scaling for the phase stiffness $\avg{W_{xy,f}^2}L\inv$ for the $f=1/64$-model at different couplings. Error bars are smaller than symbol size. Lines are guides to the eye.}%
  \label{fig:rho_fill}%
\end{figure}

\begin{figure}[tbp]
  \includegraphics{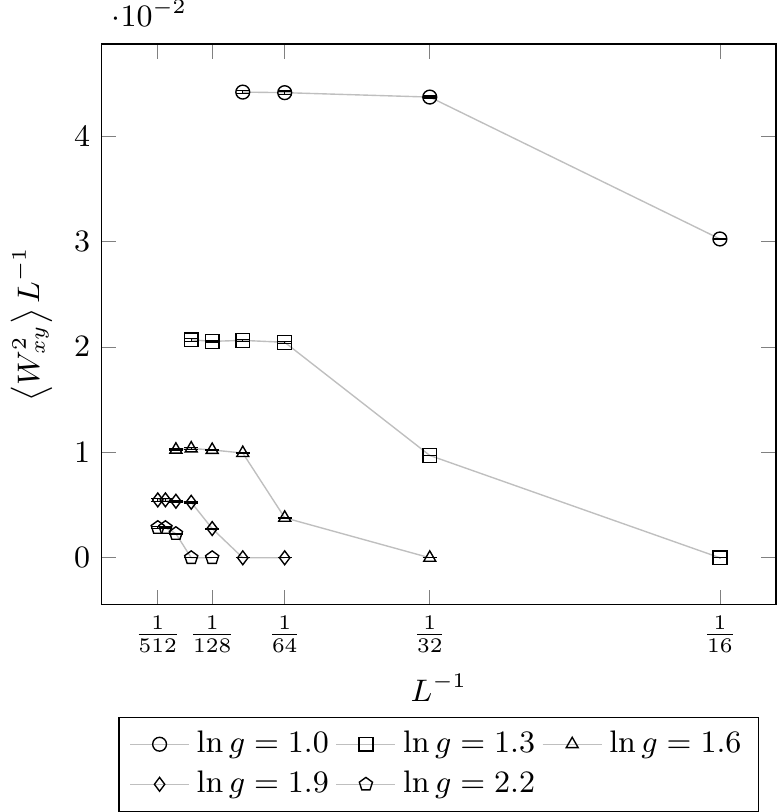}%
  \caption{Finite size scaling of $\avg{W_{xy}^2}L\inv$-curves for the LC model ($f=1$ with staggered background field) for some $\ln g$ values. $L\in \set{16,32,64,90,128,180,256,360,512}$. Lines are guides to the eye.}
  \label{fig:rho_LC_curves}
\end{figure}

\begin{figure}[tbp]
  \includegraphics{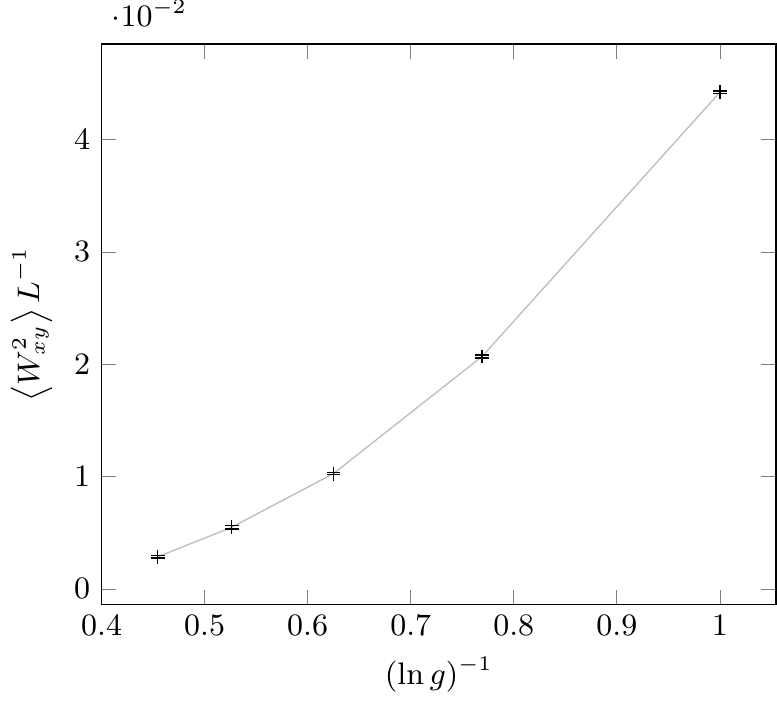}
  \caption{$\avg{W_{xy}^2}L\inv$ as a function of $(\ln g)\inv$ for the biggest system sizes simulated at the given $\ln g$ (which is assumed to be close to the value in the thermodynamic limit). $L=90$ for $(\ln g)\inv=1\inv=1$, $L=180$ for $(\ln g)\inv=1.3\inv\approx 0.77$, $L=256$ for $(\ln g)\inv=1.6\inv\approx 0.63$, and $L=512$ for $(\ln g)\inv=1.9\inv\approx 0.53$ and $(\ln g)\inv=2.2\inv\approx 0.45$. See \cref{fig:rho_LC_curves}. Lines are guides to the eye.}
  \label{fig:rho_LC_thermlim}
\end{figure}

In analogy with \cref{fig:rho_CP1}, \cref{fig:rho_fill} shows the phase stiffness scaling for the $f=1/64$-model at three selected couplings. The 
stiffness always approaches a finite value for $f>0$, indicating that there is no phase transition in the thermodynamic limit. In particular, this holds for $f=1$. 
For this case, we have explicitly shown that the behavior is essentially the same, regardless of whether $\eta_j$ is staggered or not. Phase stiffness curves for 
the LC model are shown in \cref{fig:rho_LC_curves}, while the stiffness as a function of $(\ln g)\inv$ is plotted 
in \cref{fig:rho_LC_thermlim}. Data for larger $\ln g$ (smaller $(\ln g)\inv$) values are hard to obtain given the system sizes that are required, but we find it 
reasonable to assert, based on the extensive computations presented here, that $\avg{W_{xy}^2}L\inv>0$ for $(\ln g)\inv > 0$. Note in particular how differently the
curves in \cref{fig:rho_LC_curves} behave from those in  \cref{fig:rho_CP1}. 

We conclude that for $f=0$, we obtain the phase-transition of the standard lattice $\CP$-model. In Appendix \ref{C:exponents}, we show that the critical exponents we obtain 
are consistent with those of the $2+1$-dimensional  $\groupO{3}$-model. \footnote{The exponents are in agreement with the results of Ref. \onlinecite{Takashima_PRB_2005}. 
However, the value of the critical coupling we find is roughly off by a factor of $2$. It is not clear to us what the origin of the discrepancy is.} For any nonzero value 
of $f$, the pseudocritical coupling appears to drift with system size, eventually appearing to diverge.

\section{Discussion}

Finally, we briefly summarize how to understand the suppression of the phase transition in the model defined by \cref{eq:Z_SJ}. The effect originates with 
the Berry-phase term in the action and how it appears in the LC-model, see \cref{eq:I_tau_constraint}. As noted above, in the link-current representation, 
the effect of the Berry-phase may be viewed as the introduction of a background staggered current lattice on top of which the statistically fluctuating 
link-currents are imposed, \cref{eq:I_tau_constraint}. The ordered phase of the quantum magnet is characterized by a non-zero winding number \cref{eq:stiffness}. 

A background current lattice of the form introduced by the Berry-phase facilitates the blow-out of closed current loops across the system at all coupling 
constants in the thermodynamic limit. The current lattice forms a template on which small closed current loops can connect across the system to form 
closed current loops with a linear extent scaling with the system size. This effectively represents a current-loop blowout, which is equivalent to ordering
the original spin-system. The picture is identical to the (dual) picture of type-II superconductor in a magnetic field. At zero field, there is a genuine 
phase transition from an ordered to a disordered state  driven by the proliferation of vortex loops. In a finite magnetic field, the situation is altered, and the 
field-induced vortex lattice forms a template on which small vortex-loops can connect across the system to effectively form large closed loops, thus (potentially) 
disordering the system.  

The situation where the currents loops effectively are blown out, even at couplings where only small closed current loops would exist in a zero background 
current-lattice, renders the system permanently ordered, thus suppressing the phase-transition. An alternative way of viewing it more directly in the 
spinon-gauge field description, is that the Berry phase term suppresses instanton configurations in the compact gauge-field, equivalently suppresses 
hedgehog-configurations in the action. The same result obtains also in the easy-plane limit where the $\CP$-constraint $\abs{z_1}^2+\abs{z_2}^2=1$ is 
replaced by individually constant matter-field amplitudes. In that case, the instantons that are suppressed correspond to the suppression of 
skyrmion-antiskyrmion configurations.

\appendix
\section{Exact link-current mapping of the model, \cref{eq:Z_SJ}.\label{A:current_mapping}}
We start out with a symmetrized form of the partition function,
\begin{multline}
 \mathcal{Z} = \prod_{j\mu}\int_{0}^{2\cpi}\frac{\diff A_{j\mu}}{2\cpi} \prod_{ja} \int \diff z_{ja} \diff z_{ja}^{*} \\
 \exp\left[(2g)\inv \sum_{ja\kappa} \left(z_{ja}^{*}\e{-\i A_{j\kappa}}z_{j+\mu,a} + \text{c.c.}\right) + \i 2S \sum_j \eta_j A_{j\tau} \right],
 \label{eq:Z_SJ_sym}
\end{multline}
which is obtained from \cref{eq:Z_SJ} by using that
\begin{equation}
 A_{j-\mu,\mu} = -A_{j,-\mu}.
 \label{eq:gauge_field_prop}
\end{equation}
Writing the complex scalar fields on polar form,
\begin{gather}
 z_{ja} = \rho_{ja}\e{\i\theta_{ja}}, \label{eq:z_polar} \\
 \int \diff z_{ja} \diff z_{ja}^{*} = \int_{0}^{2\cpi} \diff \theta_{ja} \int_{0}^{\infty} \diff \rho_{ja}\ \rho_{ja},
\end{gather}
the constraint \cref{eq:CP1_constraint} reads
\begin{equation}
 {\rho_{j1}}^{2} + {\rho_{j2}}^{2}  = 1 \quad \forall j.
 \label{eq:rho_constraint}
\end{equation}
We note that \cref{eq:rho_constraint} describes the unit circle arc in the first quadrant of the $\rho_{j1}\rho_{j2}$-plane (since $\rho_{ja} \geq 0$). We may therefore 
introduce a new field $\phi \in [0,\pi/2)$, given by
\begin{equation}
 (\rho_{j1}, \rho_{j2}) = (\cos \phi_{j}, \sin \phi_{j}) \label{eq:rho_phi},
\end{equation}
such that
\begin{multline}
 \left.\int_{0}^{\infty}\int_{0}^{\infty} \diff \rho_{j1}\diff \rho_{j2} \ \rho_{j1}\rho_{j2}\right|_{ {\rho_{j1}}^{2} + {\rho_{j2}}^{2} = 1 } \\
 = \int_{0}^{\frac{\cpi}{2}} \diff \phi_{j} \ \cos \phi_{j} \sin \phi_{j}.
\end{multline}
The partition function \cref{eq:Z_SJ_sym}, with the constraint \cref{eq:CP1_constraint} incorporated, can therefore be written
\begin{multline}
 \mathcal{Z} = \prod_{j\mu}\int_{0}^{2\cpi}\frac{\diff A_{j\mu}}{2\cpi} \prod_{ja} \int_{0}^{2\cpi} \diff \theta_{ja} \prod_{j} \int_{0}^{\frac{\cpi}{2}} \diff \phi_{j}\\
 \cos \phi_{j} \sin \phi_{j} \exp\left(\mathcal{S}_g + \i 2S \sum_j \eta_j A_{j\tau} \right),
 \label{eq:Z_SJ_2}
\end{multline}
where
\begin{multline}
  \mathcal{S}_g = (2g)\inv \sum_{j\kappa} \Big[\cos\phi_{j}\cos\phi_{j+\kappa}\left(\e{\i(\theta_{j+\kappa,1} - \theta_{j1} - A_{j\kappa})} + \text{c.c.}\right) \\
 \sin\phi_{j}\sin\phi_{j+\kappa}\left(\e{\i(\theta_{j+\kappa,2} - \theta_{j2} - A_{j\kappa})} + \text{c.c.}\right)\Big].
 \label{eq:S_g}
\end{multline}

Next, we split $\exp \mathcal{S}_g$ into its individual exponential factors, and Taylor expand each of them:
\begin{gather}
 \begin{split}
 \exp \mathcal{S}_g = &\prod_{j\kappa} \sum_{\substack{k_{j1\kappa} = 0\\ l_{j1\kappa} = 0}}^{\infty} \sum_{\substack{k_{j2\kappa} = 0\\ l_{j2\kappa} = 0}}^{\infty} \\
 \Bigg[ &\frac{\left((2g)\inv \cos \phi_{j} \cos \phi_{j+\kappa}\right)^{k_{j1\kappa} + l_{j1\kappa}}}{k_{j1\kappa}!l_{j1\kappa}!} \\
 &\frac{\left((2g)\inv \sin \phi_{j} \sin \phi_{j+\kappa}\right)^{k_{j2\kappa} + l_{j2\kappa}}}{k_{j2\kappa}!l_{j2\kappa}!} \\
 &\e{\i(k_{j1\kappa} - l_{j1\kappa})\left(\theta_{j+\kappa,1} - \theta_{j1} - A_{j\kappa} \right)} \\
 &\e{\i(k_{j2\kappa} - l_{j2\kappa})\left(\theta_{j+\kappa,2} - \theta_{j2} - A_{j\kappa} \right)} \Bigg].
 \end{split} 
 \label{eq:S_g_taylor_expansion}
\end{gather}
Here $k_{ja\kappa}$ and $l_{ja\kappa}$ may be seen as Taylor expansion index fields.

Inserting \cref{eq:S_g_taylor_expansion} into \cref{eq:Z_SJ_2} and rearranging (and relabelling) the terms, the partition function reads
\begin{gather}
 \begin{split}
 \mathcal{Z} = &\sum_{\set{k,l}} \prod_{j\mu}\int_{0}^{2\cpi}\frac{\diff A_{j\mu}}{2\cpi} \prod_{ja} \int_{0}^{2\cpi} \diff \theta_{ja} \prod_{j} \int_{0}^{\frac{\cpi}{2}} \diff \phi_{j} \\
 &\prod_{ja\kappa} \e{\i\theta_{ja} \left(k_{ja\kappa} - l_{ja\kappa} - k_{j+\kappa,a,-\kappa} + l_{j+\kappa,a,-\kappa}\right)}\\
 &\prod_j \e{\i 2S \eta_j A_{j\tau}} \prod_{ja\kappa} \e{\i(k_{ja\kappa} - l_{ja\kappa})A_{j\kappa}} \\
 &\prod_{j\kappa} \cos^{(k_{j1\kappa} + l_{j1\kappa} + k_{j+\kappa,1,-\kappa} + l_{j+\kappa,1,-\kappa}) + 1} \phi_j \\
 &\prod_{j\kappa} \sin^{(k_{j2\kappa} + l_{j2\kappa} + k_{j+\kappa,2,-\kappa} + l_{j+\kappa,2,-\kappa}) + 1} \phi_j \\
 & \prod_{ja\kappa}\frac{(2g)^{-(k_{ja\kappa} + l_{ja\kappa})}}{k_{ja\kappa}!l_{ja\kappa}!}.
 \label{eq:Z_SJ_3}
\end{split}
\end{gather}
$\set{k,l}$ denotes the set of all possible Taylor expansion index field configurations. 

It is convenient to introduce (what will turn out to be) the non-negative bond subcurrents
\begin{equation}
 J_{ja\kappa} \equiv k_{ja\kappa} + l_{j+\kappa,a,-\kappa} \in \mathbb{N}_0,
 \label{eq:J_current_def}
\end{equation}
as well as the total bond currents
\begin{equation}
  \quad I_{ja\kappa} \equiv J_{ja\kappa} - J_{j+\kappa,a,-\kappa} \in \mathbb{Z},
  \label{eq:I_current_def}
\end{equation}
and the factor
\begin{align}
 \mathcal{N}_{ja} &\equiv \frac{1}{2}\sum_\kappa k_{ja\kappa} + l_{ja\kappa} + k_{j+\kappa,a,-\kappa} + l_{j+\kappa,a,-\kappa} \nonumber \\
 &= \frac{1}{2} \sum_\kappa J_{ja\kappa} + J_{j+\kappa,a,-\kappa}.
 \label{eq:N_def}
\end{align}
Using these definitions, as well as \cref{eq:gauge_field_prop} and some more rearranging of terms, the partition function, \cref{eq:Z_SJ_3}, can be written 
on the form
\begin{gather}
 \begin{split}
 \mathcal{Z} = &\sum_{\set{k,l}} \prod_{j\mu}\int_{0}^{2\cpi}\frac{\diff A_{j\mu}}{2\cpi} \prod_{ja} \int_{0}^{2\cpi} \diff \theta_{ja} \prod_{j} \int_{0}^{\frac{\cpi}{2}} \diff \phi_{j} \\
 &\prod_{ja} \e{\i\theta_{ja} \sum_\kappa I_{ja\kappa}}\\
 &\prod_j \e{\i \left(2S \eta_j A_{j\tau} + \sum_\mu (I_{j1\mu} + I_{j2\mu})A_{j\mu}\right)} \\
 &\prod_{j} \cos^{2\mathcal{N}_{j1} + 1} \phi_j \sin^{2\mathcal{N}_{j2} + 1} \phi_j\\
 & \prod_{ja\kappa}\frac{(2g)^{-J_{ja\kappa}}}{k_{ja\kappa}!l_{ja\kappa}!}.
 \label{eq:Z_SJ_4}
\end{split}
\end{gather}
Note that the summation goes over \emph{positive} directions only in the gauge field factor.

The partition function is now on a form where the integrals are decoupled and may be performed easily.

The integration of the $\theta$-field in \cref{eq:Z_SJ_4} gives just a Kronecker delta (up to an irrelevant scaling factor) at each lattice site. Hence, we obtain the total 
bond-current conservation constraint, or \enquote{Kirchhoff's law},
\begin{equation}
 \sum_\kappa I_{ja\kappa} = 0, \quad \forall j,a.
 \label{eq:I_app_kirch}
\end{equation}
Note also that this, by the definition \cref{eq:I_current_def}, implies that \cref{eq:N_def} may be simplified to
\begin{equation}
 \mathcal{N}_{ja} = \sum_\kappa J_{ja\kappa} \in \mathcal{N}_0.
\end{equation}

In the same way as for the $\theta$-integration, the gauge-field integration gives Kronecker deltas, leading to the coupling of the components,
\begin{equation}
\begin{gathered}
 I_{j1x} + I_{j2x} = 0, \\
 I_{j1y} + I_{j2y} = 0, \\
 I_{j1\tau} + I_{j2\tau} + 2S\eta_j = 0.
\end{gathered} 
\label{eq:I_app_const}
\end{equation}

The $\phi$-field is integrated out by
\begin{multline}
 \int_{0}^{\frac{\cpi}{2}} \diff \phi_{j} \cos^{2\mathcal{N}_{j1} + 1}\phi_{j}\sin^{2\mathcal{N}_{j2} + 1}\phi_{j} \\
 = \frac{\mathcal{N}_{j1}!\mathcal{N}_{j2}!}{2\left(\mathcal{N}_{j1} + \mathcal{N}_{j2} + 1\right)!},
 \label{eq:phi_integral}
\end{multline}
where we have used the identity
\begin{multline}
 \int_{0}^{\frac{\cpi}{2}} \cos^{m}x \sin^{n} x \diff x \\
 = \frac{\Gamma\left(\frac{m+1}{2}\right)\Gamma\left(\frac{n+1}{2}\right)}{2\Gamma\left(\frac{m+n+2}{2}\right)}, \quad m,n > 0.
 \label{eq:intcossin}
\end{multline}
We will ignore the physically irrelevant multiplicative factor of $2$ in the denominator of \cref{eq:phi_integral} in the final expression for the 
partition function.

The last factor we have to deal with is
\begin{equation}
 \sum_{\set{k,l}}\prod_{ja\kappa}\frac{(2g)^{-J_{ja\kappa}}}{k_{ja\kappa}!l_{ja\kappa}!}.
 \label{eq:index_field_part}
\end{equation}
We have included the sum over all possible $k,l$ field configurations, as we want to change \cref{eq:index_field_part} to a sum over all possible $J$-current field 
configurations, $\set{J}$, instead. (There is no problem with this, as all the other terms in the partition function and the constraints are -- as we have seen -- 
exclusively $J$ dependent.) Using the definition of the positive bond currents, \cref{eq:J_current_def}, as well as some standard combinatorial results, we may 
rewrite \cref{eq:index_field_part} as
\begin{align}
 \sum_{\set{k,l}}\prod_{ja\kappa} \frac{(2g)^{-J_{ja\kappa}}}{k_{ja\kappa}! l_{ja\kappa}!} 
 &= \sum_{\set{J}}\prod_{ja\kappa} \sum_{k_{ja\kappa} = 0}^{J_{ja\kappa}}\frac{(2g)^{-J_{ja\kappa}}}{k_{ja\kappa}! (J_{ja\kappa} - k_{ja\kappa})!}  \nonumber \\
 &= \sum_{\set{J}}\prod_{ja\kappa} \frac{(2g)^{-J_{ja\kappa}}}{J_{ja\kappa}!} \sum_{k_{ja\kappa} = 0}^{J_{ja\kappa}} \binom{J_{ja\kappa}}{k_{ja\kappa}} \nonumber \\
 &= \sum_{\set{J}}\prod_{ja\kappa} \frac{(2g)^{-J_{ja\kappa}}}{J_{ja\kappa}!}2^{J_{ja\kappa}} \nonumber \\
 &= \sum_{\set{J}}\prod_{ja\kappa} \frac{g^{-J_{ja\kappa}}}{J_{ja\kappa}!}
 \label{eq:index_field_2}
\end{align}

Collecting all of the above gives the desired results \cref{eq:Z_current,eq:kirchhoff,eq:I_x_constraint,eq:I_y_constraint,eq:I_tau_constraint}.

\section{Link-current representation of the model with basic representation of the Berry-phase \label{B:Berry}}
The basic form of the Berry-phase contribution to the action is given by \cite{Read_Sachdev_PRB_1990}
\begin{equation}
 \mathcal{S}_{\text{B}} = 2S\sum_{ia}\eta_i \int_{0}^\beta  z_{a}^*(\v r_{i},\tau) \deriv{z_{a}(\v r_{i},\tau)}{\tau} \diff \tau,
\end{equation}
which, when discretizing imaginary time and ignoring irrelevant constants, may be written
\begin{gather}
 \begin{split}
  \mathcal{S}_{\text{B}} &= 2S\sum_{ja}\eta_j  z_{ja\tau}^* z_{j+\tau,a\tau}\\
 &= 2S\sum_{j}\eta_{j}\Bigl(\cos \phi_{j} \cos \phi_{j+\tau}\e{\i(\theta_{j+\tau,1}-\theta_{j1})} \\
 &\hphantom{{}=2S\sum_{j}\eta_{j}\Bigl(} + \sin \phi_{j} \sin \phi_{j+\tau}\e{\i(\theta_{j+\tau,2}-\theta_{j2})}\Bigr),
 \label{eq:S_B_true}
 \end{split}
\end{gather}
in the $\beta \to \infty$ limit. We have introduced the fields $\theta$ and $\phi$ defined in \cref{eq:z_polar,eq:rho_phi}. Note that \cref{eq:S_B_true} may 
\emph{not} be symmetrized.

Replacing the Berry-phase term of partition function \eqref{eq:Z_SJ_2}, $\i 2S \sum_j \eta_j A_{j\tau}$, with \cref{eq:S_B_true}, the link-current mapping 
may proceed in the same way as in Appendix \ref{A:current_mapping}. The details of how to approximate the form of the Berry phase given in \cref{eq:S_B_true} to
the form of the Berry phase given in \cref{eq:Z_SJ}, is provided in Chapter 13 of Ref. \onlinecite{Sachdev_Book}. 

Taylor expanding $\exp \mathcal{S}_{\text{B}}$ gives an additional expansion index field $m_{ja} \in \mathbb{N}_{0}$ coupling in the $\tau$ direction, so the partition 
function equivalent to \cref{eq:Z_SJ_4} reads
\begin{gather}
 \begin{split}
 \mathcal{Z} = &\sum_{\set{k,l,m}} \prod_{j\mu}\int_{0}^{2\cpi}\frac{\diff A_{j\mu}}{2\cpi} \prod_{ja} \int_{0}^{2\cpi} \diff \theta_{ja} \prod_{j} \int_{0}^{\frac{\cpi}{2}} \diff \phi_{j} \\
 &\prod_{ja} \e{\i\theta_{ja} \left(m_{j-\tau,a} - m_{j,a} + \sum_\kappa I_{ja\kappa} \right)}\\
 &\prod_j \e{\i \sum_\mu (I_{j1\mu} + I_{j2\mu})A_{j\mu}} \\
 &\prod_{j} \cos^{2\mathcal{N}_{j1} + 1 + m_{j1} + m_{j-\tau,1} } \phi_j \sin^{2\mathcal{N}_{j2} + 1 +  m_{j2} + m_{j-\tau,2}} \phi_j\\
 & \prod_{ja\kappa}\frac{(2g)^{-J_{ja\kappa}}}{k_{ja\kappa}!l_{ja\kappa}!}\prod_{ja}\frac{(2S\eta_j)^{m_{ja}}}{m_{ja}!}.
\end{split}
\end{gather}
The field integrals can be done as before, leading to the new constraints
\begin{gather}
 m_{j-\tau,a} - m_{j,a} + \sum_\kappa I_{ja\kappa} = 0, \\
 I_{j1\mu} + I_{j2\mu} = 0.
\end{gather}
Using \cref{eq:index_field_2} (which is independent of the $m$-field, and thus still valid) and \cref{eq:intcossin}, we are left with the partition function
\begin{multline}
 \mathcal{Z} = \sum_{\set{J,m}} \prod_{ja}\frac{(2S\eta_j)^{m_{ja}}}{m_{ja}!}\prod_{ja\kappa}\frac{g^{-J_{ja\kappa}}}{J_{ja\kappa}!} \\
 \prod_{j}\frac{\prod_{a}\Gamma \left(\mathcal{N}_{ja} + 1 + \frac{1}{2}(m_{ja} + m_{j-\tau,a}) \right) }{\Gamma\left(\sum_{a}\mathcal{N}_{ja} + 1 + \frac{1}{2}(m_{ja} + m_{j-\tau,a})\right)},
\end{multline}
which should be compared to \cref{eq:Z_current}. The main problem with this formulation is that a sign problem arises from the factors $\eta_{j}^{m_{ja}} = \pm 1$, rendering 
the model hard to deal with in Monte Carlo simulations.

\section{Critical exponents for the $\CP$-model  \label{C:exponents}}
If we define the global magnetization $\v{m} \equiv N^{-1}\sum_j \v n_j, \quad N \equiv L^3$, and use the definition of $\v{n}_j$ in terms of Pauli matrices and the $z$-fields, \cref{eq:z_definition}, we end up with the relation
\begin{equation}
 \avg{m_{x}^2 + m_{y}^2} = \frac{2}{N}\sum_j h(j),
\end{equation}
where
\begin{equation}
 h(i-j) \equiv \avg{z_{i1}z_{j1}^*z_{i2}^*z_{j2} + \text{c.c.}}.
 \label{eq:corr}
\end{equation}
Now, for the $\CP$-model, we have
\begin{equation}
  \avg{z_{i'1}z_{j'1}^*z_{i'2}^*z_{j'2}} = \mathcal{Z}_{\CP}\inv \mathcal{W}_{i'j'},
\end{equation}
where
\begin{multline}
  \mathcal{W}_{i'j'} = \prod_{j\mu}\int_{0}^{2\cpi}\frac{\diff A_{j\mu}}{2\cpi} \prod_{ja} \int \diff z_{ja} \diff z_{ja}^{*}  z_{i'1}z_{j'1}^*z_{i'2}^*z_{j'2} \\
 \exp\left[g\inv \sum_{ja\mu} \left(z_{ja}^{*}\e{-\i A_{j\mu}}z_{j+\mu,a} + \text{c.c.}\right)\right],
 \label{eq:Wij}
\end{multline}
with the usual $\CP$-constraint \cref{eq:CP1_constraint}. In the link-current formalism, \cref{eq:Wij} becomes (using the procedures of \cref{A:current_mapping})
\begin{gather}
 \mathcal{W}_{i'j'} = \sum_{\set{J}} \prod_{ja\kappa}\frac{g^{-J_{ja\kappa}}}{J_{ja\kappa}!} \prod_{j\neq i',j'} \left[ \frac{\mathcal{N}_{j1}!\mathcal{N}_{j2}!}{\left(\mathcal{N}_{j1} + \mathcal{N}_{j2} + 1\right)!}\right] Q_{i'j'}, \\
 Q_{i'j'} = \left\{
 \begin{aligned}
  &\frac{\left(\mathcal{N}_{i'1} + 1\right)!\left(\mathcal{N}_{i'2} + 1\right)!}{\left(\mathcal{N}_{i'1} + \mathcal{N}_{i'2} + 3\right)!},\quad i'=j' \\
  &\frac{\prod_a \Gamma\left(\mathcal{N}_{i'a} + \frac{3}{2}\right)\Gamma\left(\mathcal{N}_{j'a} + \frac{3}{2}\right)  }{\left(\mathcal{N}_{i'1} + \mathcal{N}_{i'2} + 2\right)!\left(\mathcal{N}_{j'1} + \mathcal{N}_{j'2} + 2\right)!},  \quad i'\neq j',
 \end{aligned}
  \right.
\end{gather}
with the constraints
\begin{gather}
 I_{j1\mu} + I_{j2\mu} = 0, \\
 \sum_{\kappa} I_{j1\kappa} = \left\{
 \begin{aligned}
  -1,& \quad j = i'\\
  1,& \quad j = j' \\
  0,&\quad j \neq i',j'
 \end{aligned}
 \right.\label{eq:open_worm_constraint}
\end{gather}
We get $\mathcal{W}_{i',j'}^*$ by interchanging $i' \leftrightarrow j'$ in the last constraint.

\Cref{eq:open_worm_constraint} means that $h(i-j)$ must be sampled for a field configuration where all current loops/worms \emph{but one} are closed. The open worm has its head(tail) at lattice site $i$ and tail(head) at $j$. Since the worm is already following the probability distribution given by the partition function $\mathcal{Z}_{\CP}$, the weight associated with this \enquote{background} distribution must be divided out before we can sample $h$ properly. Hence, to sample $\avg{m_{xy}^2} \equiv \avg{m_{x}^2 + m_{y}^2}/2$, we store
\begin{equation}
 M^2 \gets M^2 + \left\{\begin{aligned}
  &\frac{\left(\mathcal{N}_{i1} + 1\right)\left(\mathcal{N}_{i2} + 1\right)}{\left(\mathcal{N}_{i1} + \mathcal{N}_{i2} + 2\right)\left(\mathcal{N}_{i1} + \mathcal{N}_{i2} + 3\right)}, \quad i-j = 0\\
  &\frac{1}{\left(\mathcal{N}_{i1} + \mathcal{N}_{i2} + 2\right)\left(\mathcal{N}_{j1} + \mathcal{N}_{j2} + 2\right)} \\
  & \times \prod_{a}\frac{\Gamma\left(\mathcal{N}_{ia} + \frac{3}{2}\right)\Gamma\left(\mathcal{N}_{ja} + \frac{3}{2}\right)}{\mathcal{N}_{ia}!\mathcal{N}_{ja}!}, \quad i-j \neq 0
 \end{aligned}
 \right.
\end{equation}
at each Monte Carlo step, along with
\begin{equation}
 Z \gets Z + 1
\end{equation}
each time the worm closes ($i = j$). The unbiased Monte Carlo estimator $\avg{m_{xy}^2}_{\text{MC}}$ is then given by
\begin{equation}
 \avg{m_{xy}^2}_{\text{MC}} = \frac{M^2}{NZ}
\end{equation}

\begin{figure}[tbp]
  \includegraphics{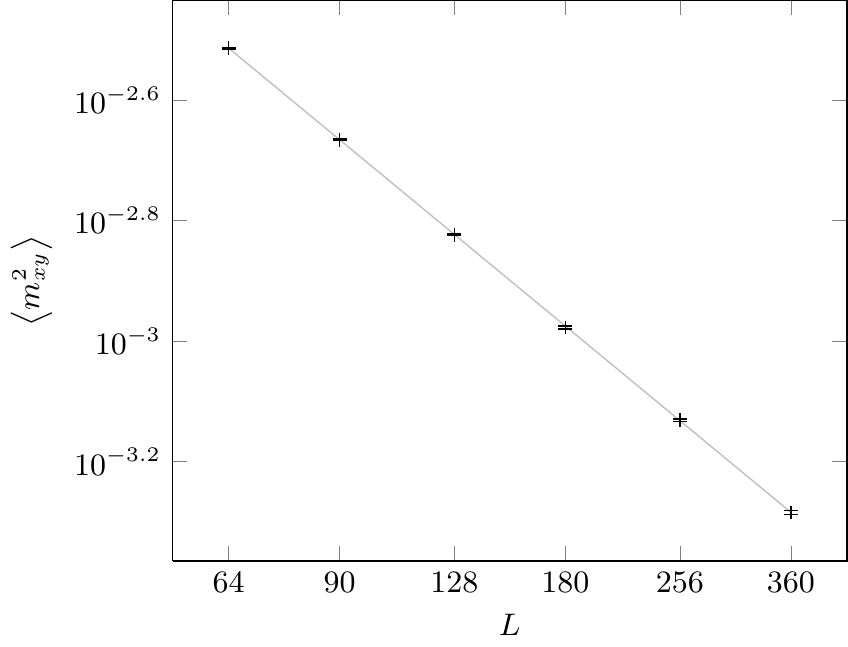}%
  \caption{log-log plot of the finite size scaling of $\avg{m_{xy}^2}$ for the $\CP$-model at $\ln g = \num{0.414504} \approx \ln g_\text{c}$ (markers), plotted together with a scaling curve $\sim L^{-2 \times 0.513}$ (light gray).}
  \label{fig:m2_CP1_scaling}
\end{figure}

From FSS, we expect $\avg{m_{xy}^2} \sim L^{-\frac{2\beta}{\nu}}$ at criticality. We find $\beta/\nu = \num{0.513 +- 0.004}$ in an FSS analysis for system sizes up to $L = 360$, see \cref{fig:m2_CP1_scaling}.\footnote{The FSS simulations were performed at $\ln g = 0.414504$, without any reweighting involved (due to technicalities in the simulations). After this result was obtained, it was found that $\ln g = \num{0.414508}$ probably is closer to $\ln g_{\text{c}}$. However, the difference is smaller than the statistical fluctuations of the simulations, so redoing them was deemed unnecessary.} This is in reasonable agreement with the $\groupO{3}$ universality class result of Ref. \onlinecite{Campostrini_et_al_PRB_2002}, 
$\beta/\nu = \num{0.5187+-0.0006}$.

\begin{acknowledgments}
TAB thanks NTNU for financial support and the Norwegian consortium for high-performance computing (NOTUR) for computer time and technical support. AS was supported by the Research Council of Norway, through Grants 205591/V20 and 216700/F20. 
\end{acknowledgments}

\bibliography{references}

\end{document}